\documentclass[twoside,twocolumn,9pt]{article}
\usepackage{extsizes}
\usepackage[super,sort&compress,comma]{natbib} 
\usepackage[version=3]{mhchem}
\usepackage[left=1.5cm, right=1.5cm, top=1.785cm, bottom=2.0cm]{geometry}
\usepackage{balance}
\usepackage{mathptmx}
\usepackage{sectsty}
\usepackage{graphicx} 
\usepackage{lastpage}
\usepackage[format=plain,justification=justified,singlelinecheck=false,font={stretch=1.125,small,sf},labelfont=bf,labelsep=space]{caption}
\usepackage{float}
\usepackage{fancyhdr}
\usepackage{fnpos}
\usepackage[english]{babel}
\addto{\captionsenglish}{%
  
}
\usepackage{array}
\usepackage{droidsans}
\usepackage{charter}
\usepackage[T1]{fontenc}
\usepackage[usenames,dvipsnames]{xcolor}
\usepackage{setspace}
\usepackage[compact]{titlesec}
\usepackage{hyperref}
\usepackage{gensymb} 

\definecolor{cream}{RGB}{222,217,201}

\begin{document}

\pagestyle{fancy}
\thispagestyle{plain}
\fancypagestyle{plain}{
\renewcommand{\headrulewidth}{0pt}
}

\makeFNbottom
\makeatletter
\renewcommand\LARGE{\@setfontsize\LARGE{15pt}{17}}
\renewcommand\Large{\@setfontsize\Large{12pt}{14}}
\renewcommand\large{\@setfontsize\large{10pt}{12}}
\renewcommand\footnotesize{\@setfontsize\footnotesize{7pt}{10}}
\renewcommand\scriptsize{\@setfontsize\scriptsize{7pt}{7}}
\makeatother

\renewcommand{\thefootnote}{\fnsymbol{footnote}}
\renewcommand\footnoterule{\vspace*{1pt}%
\color{cream}\hrule width 3.5in height 0.4pt \color{black} \vspace*{5pt}} 
\setcounter{secnumdepth}{5}

\makeatletter 
\renewcommand\@biblabel[1]{#1}            
\renewcommand\@makefntext[1]%
{\noindent\makebox[0pt][r]{\@thefnmark\,}#1}
\makeatother 
\renewcommand{\figurename}{\small{Fig.}~}
\sectionfont{\sffamily\Large}
\subsectionfont{\normalsize}
\subsubsectionfont{\bf}
\setstretch{1.125} 
\setlength{\skip\footins}{0.8cm}
\setlength{\footnotesep}{0.25cm}
\setlength{\jot}{10pt}
\titlespacing*{\section}{0pt}{4pt}{4pt}
\titlespacing*{\subsection}{0pt}{15pt}{1pt}

\fancyfoot{}
\fancyfoot[LO,RE]{\vspace{-7.1pt}\includegraphics[height=9pt]{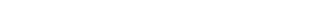}}
\fancyfoot[CO]{\vspace{-7.1pt}\hspace{13.2cm}\includegraphics{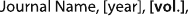}}
\fancyfoot[CE]{\vspace{-7.2pt}\hspace{-14.2cm}\includegraphics{RF}}
\fancyfoot[RO]{\footnotesize{\sffamily{1--\pageref{LastPage} ~\textbar  \hspace{2pt}\thepage}}}
\fancyfoot[LE]{\footnotesize{\sffamily{\thepage~\textbar\hspace{3.45cm} 1--\pageref{LastPage}}}}
\fancyhead{}
\renewcommand{\headrulewidth}{0pt} 
\renewcommand{\footrulewidth}{0pt}
\setlength{\arrayrulewidth}{1pt}
\setlength{\columnsep}{6.5mm}
\setlength\bibsep{1pt}

\makeatletter 
\newlength{\figrulesep} 
\setlength{\figrulesep}{0.5\textfloatsep} 

\newcommand{\topfigrule}{\vspace*{-1pt}%
\noindent{\color{cream}\rule[-\figrulesep]{\columnwidth}{1.5pt}} }

\newcommand{\botfigrule}{\vspace*{-2pt}%
\noindent{\color{cream}\rule[\figrulesep]{\columnwidth}{1.5pt}} }

\newcommand{\dblfigrule}{\vspace*{-1pt}%
\noindent{\color{cream}\rule[-\figrulesep]{\textwidth}{1.5pt}} }

\makeatother

\twocolumn[
  \begin{@twocolumnfalse}
{\includegraphics[height=30pt]{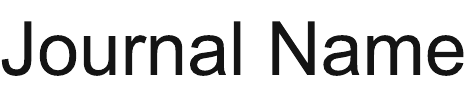}\hfill\raisebox{0pt}[0pt][0pt]{\includegraphics[height=55pt]{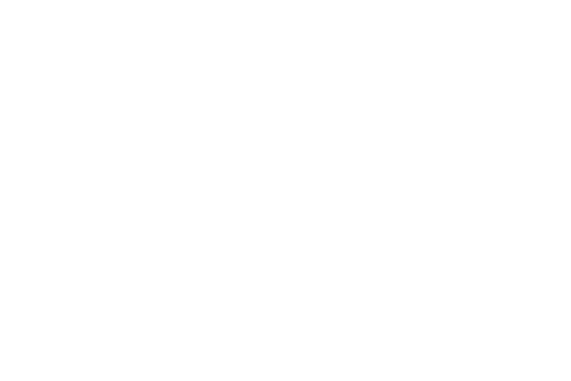}}\\[1ex]
\includegraphics[width=18.5cm]{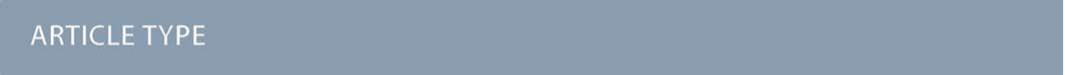}}\par
\vspace{1em}
\sffamily
\begin{tabular}{m{4.5cm} p{13.5cm} }

\includegraphics{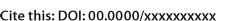} & \noindent\LARGE{\textbf{The motion of catalytically active colloids approaching a surface
$^\dag$}} \\
 & \vspace{0.3cm} \\

 & \noindent\large{Julio Melio,$^{\ast}$\textit{$^{a}$} Solenn Riedel,$^{\ast}$\textit{$^{a}$} Ali Azadbakht\textit{$^{a}$}, Silvana A. Caipa Cure\textit{$^{a}$}, Tom M.J. Evers\textit{$^{b}$}, Mehrad Babaei\textit{$^{b}$}, Alireza Mashaghi\textit{$^{b}$}, Joost de Graaf,\textit{$^{c}$} and Daniela J. Kraft\textit{$^{a}$}} \\

\includegraphics{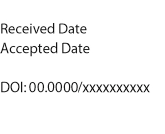} & \noindent\normalsize{Catalytic microswimmers typically swim close to walls due to hydrodynamic and/or phoretic effects. The walls in turn are known to affect their propulsion, making it difficult to single out the contributions that stem from particle-based catalytic propulsion only, thereby preventing an understanding of the propulsion mechanism. Here, we use acoustic tweezers to lift catalytically active Janus spheres away from the wall to study their motion in bulk and when approaching a wall. Mean-squared displacement analysis shows that diffusion constants at different heights match with Faxén's prediction for the near-wall hydrodynamic mobility. Both particles close to a substrate and in bulk show a decrease in velocity with increasing salt concentration, suggesting that the dominant factor for the decrease in speed is a reduction of the swimmer-based propulsion. The velocity-height profile follows a hydrodynamic scaling relation as well, implying a coupling between the wall and the swimming speed. The observed speed reduction upon addition of salt matches expectations from a electrokinetic theory, except for experiments in 0.1 wt\% \ce{H2O2} in bulk, which could indicate contributions from a different propulsion mechanism. Our results help with the understanding of ionic effects on microswimmers in 3D and point to a coupling between the wall and the particle that affects its self-propulsion speed.}

\end{tabular}

 \end{@twocolumnfalse} \vspace{0.6cm}

  ]

\renewcommand*\rmdefault{bch}\normalfont\upshape
\rmfamily
\section*{}
\vspace{-1cm}


\footnotetext{\textit{$^{a}$~Huygens-Kamerlingh Onnes Laboratory, Leiden University, P.O. Box 9504, 2300 RA Leiden, The Netherlands}}
\footnotetext{\textit{$^{b}$~Medical Systems Biophysics and Bioengineering, Leiden Academic Centre for Drug Research, Faculty of Science, Leiden University, 2333CC, Leiden, The Netherlands}}
\footnotetext{\textit{$^{c}$~Institute for Theoretical Physics, Center for Extreme Matter and Emergent Phenomena, Utrecht University, Princetonplein 5, 3584 CC, Utrecht, The Netherlands}}

\footnotetext{\dag~Electronic Supplementary Information (ESI) available: [details of any supplementary information available should be included here]. See DOI: 00.0000/00000000.}

\footnotetext{$^\ast$~ These authors contributed equally to this work.}




\rmfamily 


\begin{figure}[t]
\centering
  \includegraphics[width=0.5\textwidth]{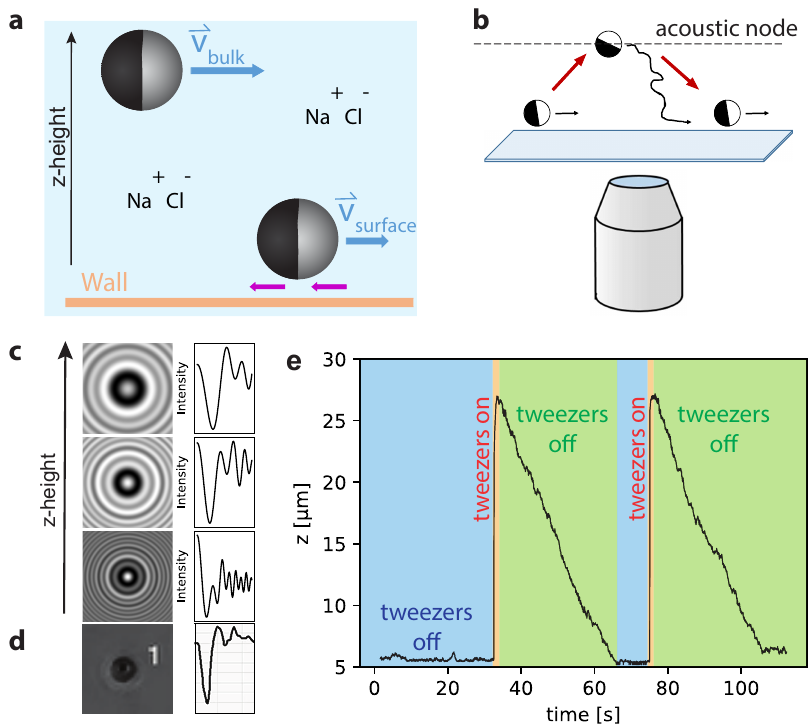}
  \caption{\textit{Acoustic tweezer experiments.} (a) Close to the substrate, active particles significantly slow down upon the addition of salt. By observing self-propelled particles in bulk, we can determine whether these salt effects are wall-effects or if salt also affects the bulk behavior. The purple arrows represent the counter flows that occur near the wall. (b) Schematic drawing of the acoustic tweezers setup with acoustic node. (c) Three exemplary predictions of scattering pattern for a polystyrene particle at different z-heights with the corresponding radial intensity profile. This information is needed to connect the holographic signal to the distance from the focal plane. (d) Experimental scattering pattern snapshot with the radial intensity profile for a particle $\approx10$~$\mu$m above the substrate. (e) $z$-Coordinate trajectory for a passive particle that is lifted up with the acoustic tweezers and sediments twice.}
  \label{fig1}
\end{figure}
Catalytic synthetic microswimmers~\cite{Howse2007, Palacci2013, Paxton2004} are great model systems for living active agents like motile bacteria~\cite{Berke2008, Peruani2012, Petroff2015}, algae~\cite{Kantsler2013}, and sperms~\cite{Rothschild1963}. When dispersed in a fuel solution these colloidal particles self-propel as a result of solute gradients generated by the asymmetric catalytic decomposition of the fuel on the swimmer's surface~\cite{Golestanian2005, Brady2011, Ebbens2014, Brown2014, Takagi2014, Dey2015, Safdar2015}. In experiments, synthetic active particles typically have affinity for surfaces, resulting in them moving close to substrates. The motility of their biological counterparts close to boundaries, however, is very different from bulk motility~\cite{Lauga2009, Bianchi2017, Berke2008, Rothschild1963, Bohmer2005, Lauga2006}. Similarly, also for synthetic swimmers the properties of a nearby wall were shown to have a significant impact on e.g. the propulsion speed of the particle~\cite{Ketzetzi2020, Wei2018, Holterhoff2018, Heidari2020} and the swimmer-wall separation~\cite{Ketzetzi2020a}, suggesting the presence of a hydrodynamic and/or phoretic coupling with the wall. Additionally, previous work from our group has suggested that the interaction of the wall with the chemical species generated by the catalytic microswimmer can lead to the occurrence of counter flows near the substrate that could affect the net velocity of the swimmer~\cite{Ketzetzi2020, Ketzetzi2020a}.
\\
\\
More clues about potential wall-particle interactions stem from experiments with salt. The speed of swimmers near substrates has been shown to drastically reduce upon the addition of sodium chloride or other charged species~\cite{Ketzetzi2020a, Ebbens2014, Brown2014}. This observation provides interesting insights into the nature of the propulsion mechanism of active colloids. The mechanism itself remains under discussion and might have contributions from momentum transfer, as well as neutral and/or ionic self-diffusiophoresis and electrophoresis~\cite{Brown2014, Brown2017, Eloul2020, Ketzetzi2020, Ketzetzi2020a}.
The near-substrate observations of ion induced slowing down suggests an electrophoresis-based propulsion mechanism\cite{Brown2014,Brown2017}, or  could result from a reduction of ionic near-wall counter flows.\cite{Ketzetzi2020a} The latter proposal would reconcile observations of a lack of speed variation with the zeta potential of the bare particles, as well as those of a constant swimming height from the substrate at various salt concentrations~\cite{Ketzetzi2020a}. This would make the presence of a wall crucial to observe this salt-induced slowing down, and conversely suggest that the speed of the same particles in bulk, far away from the wall, would be unaffected by the addition of salt. 
\\
\\
Motivated by these different proposals, we here compare speeds in bulk and close to a substrate to isolate a possible contribution from the osmotic flows generated on the substrate (cf. Figure ~\ref{fig1}a). To do so, we performed measurements on active particles that were lifted into the bulk solution with the help of acoustic tweezers, see Figure \ref{fig1}b, and compared their behavior at different salt concentrations. Bulk experiments were performed at heights ranging from 10 to 30~$\mu$m above the substrate, which is greater than, and at the smallest separations comparable to, the decay length of the phoretic flows around the particle~\cite{Campbell2019}. We observed a decrease in velocity with increasing ionic concentration which is similar for particles in bulk and close to the substrate. This suggests that adding ions primarily affects the particle-based contributions to the propulsion. If a significant osmotic flow along the wall would be present and affect the motion of the active particles, then it only appears to have a minor effect on the self-propulsion speed. 
\\
\\ 
\section*{Experimental}
\textbf{Experimental setup.} In all experiments, we used (4.50~$\pm$~0.14)~$\mu$m sized polystyrene (PS) particles half-coated with a 5~nm thick Pt/Pd (80/20) layer.\footnote[1]{The metal layer is applied on one side, leading to a maximum radial thickness of 5~nm at a pole that gradually decreases to 0 at the equator.} These were rendered active by dispersing them in a 0.1~wt\% or 0.5~wt\% aqueous hydrogen peroxide (\ce{H2O2}) solution. In this system of catalytic synthetic microswimmers, self-propulsion is driven by solute gradients generated through the catalytic decomposition of \ce{H2O2} at the  Pt/Pd cap~\cite{Howse2007, Ebbens2014}. The active particles were lifted into the bulk using acoustic tweezers paired with holographic microscopy. The measurements were performed in a G2 AFS microfluidic chip holder using an AFS-G2 acoustic tweezers setup from Lumicks B.V. with a motorized $z$-stage mounted on an inverted microscope with a Nikon $20\times$ air objective. For bulk measurements, the particles were lifted up to the acoustic node $\pm~20~\mu$m above the substrate using a standing acoustic wave. When the acoustic field is switched off, the particles were free to self-propel in 3D and the particle positions were recorded in all three dimensions. 
\\
\\
\textbf{Extraction of particle position.} To obtain a particle's $z$-coordinate, a look-up-table (LUT) with a step-size of 100~nm was produced for each particle in the field of view before generating the acoustic wave (Fig. \ref{fig1}d). The LUT comprises the scattering patterns of the particle at specific distances from the focal plane. This allows the radial scattering patterns recorded for that same active particle far from the substrate to be translated to a particle height. The presence of the metal cap influences these radial scattering patterns based on the particle orientation. However, we neglect this effect and treat our particles as isotropic. We use the small cap thickness and the fact that an individual LUT is recorded for each particle before it is lifted to justify this approximation. For illustration purposes, examples of scattering patterns of PS-particles at different $z$-heights and the corresponding radial intensity profiles, which were calculated using the python package Holopy, are shown in Figure~\ref{fig1}c. The $x$-, $y$-, and $z$-coordinates are obtained by particle tracking methods using the LabVIEW software provided by Lumicks B.V. Figure~\ref{fig1}e shows a typical trajectory of an experiment with a passive particle in water. After sedimentation in the microfluidic cell, the particle first is close to the substrate (blue section). Upon switching on the acoustic tweezers, the particle is lifted to the acoustic node (orange section), from where it sediments back to the bottom of the cell when the tweezers are switched off (green section). Active particles were observed to move both downwards as well as upwards in the experimental cell after being released from the acoustic trap~\cite{Carrasco-Fadanelli2023}. 2D experiments close to the substrate were performed after sedimentation of the particles whilst the acoustic tweezers were off. Experiments on the substrate were complemented by additional measurements using an inverted microscope (Nikon Ti-E) equipped with a 60$\times$ water immersion objective (NA~$=$~1.2). The particle motion was imaged at a framerate of 20 fps.
\\
\\
\section*{Results and Discussion}
\begin{figure}
 \centering
 \includegraphics[width=0.5\textwidth]{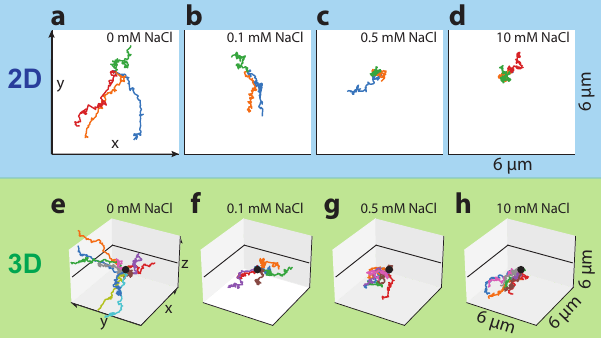}
 \caption{\textit{Influence of salt on the activity of catalytic microswimmers in both 2D and 3D.} Trajectories of active particles for 2D experiments (a-d) and 3D experiments (e-h). All trajectories are plotted for 100 consecutive frames (with a frame rate of 18.9~fps corresponding to about 30~s) and have been moved to start in the same point.}
 \label{fig2}
\end{figure}

\begin{figure}[t!]
\centering
  \includegraphics[width=0.5\textwidth]{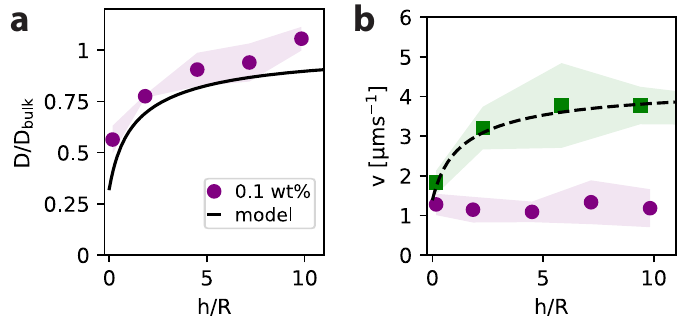}
  \caption{\textit{Height dependence of the diffusion coefficient and particle velocity.} (a) For a single particle suspended in 0.1~wt\% \ce{H2O2}, the diffusion coefficient $D$ normalized over the expectation value for $D_\mathrm{bulk}$ depends on the particle height above the substrate $h$ normalized over the particle radius $R$. The evolution of $D(h)$ over $D_\mathrm{bulk}$ follows the hydrodynamic model for the height-dependent diffusion derived by Faxén~\cite{Hilding1921}. (b) The single particle velocity $v$, however, is independent of the distance between particle and substrate for low activity (0.1~wt\% \ce{H2O2}), but becomes height-sensitive at higher activity (0.5~wt\% \ce{H2O2}), when the particle is closer than approx.~5 particle radii above the substrate. The dotted line in (b) serves to guide the eye. Shaded regions in all plots indicate standard deviation between different particles under the same conditions. All shown experiments are in absence of salt.}
  \label{fig3}
\end{figure}

\begin{figure}[t!]
\centering
  \includegraphics[width=0.5\textwidth]{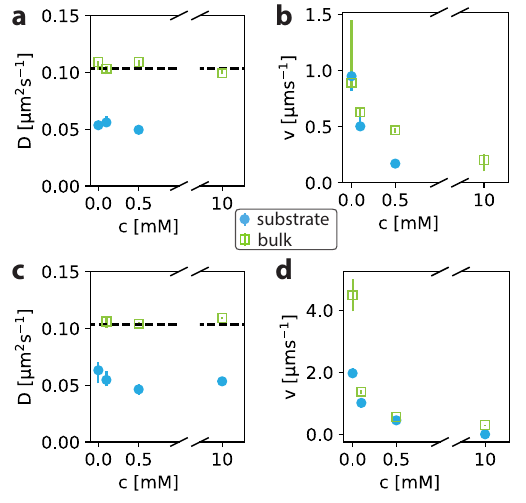}
  \caption{\textit{Salt affects active particles close to the surface and in bulk in a similar way. } Diffusion coefficient $D$ and particle velocity $v$ both as a function of salt concentration for particles suspended in 0.1~wt\% \ce{H2O2} (a and b) and 0.5~wt\% \ce{H2O2} (c and d), respectively. Data obtained close to the substrate is indicated by blue spheres, and from the bulk by green squares. $D$ and $v$ were obtained from a mean-squared displacement (MSD) analysis where lag times up to 1~s were considered. The dashed line in a and c indicates the expectation value from the Stokes-Einstein relation with $T=296$~K and $\eta=0.9321$~mPas (water)~\cite{Huber2009}. Points plotted are median values and error bars indicate the first and third quartiles.}
  \label{fig4}
\end{figure}

\textbf{Salt addition experiments.}
To compare the effect of salt on the particle motion in bulk and close to the substrate, we performed a series of experiments at different \ce{NaCl} concentrations ($c_\text{NaCl}$). 2D and 3D measurements, with the acoustic tweezers off or on, respectively, were taken at $c_\text{NaCl} = 0$, 0.1, 0.5, and 10~mM in an aqueous \ce{H2O2} solution with a concentration of 0.1~wt\% or 0.5~wt\%. Exemplary associated 30~s trajectories for increasing salt concentrations are shown in Fig.~\ref{fig2}, where different colors indicate different particle trajectories. Panels a-d correspond to colloids self-propelling at the bottom of the measurement cell and thus close to a wall and hence are shown as 2D plots. Panels e-h correspond to colloids moving in 3D far away from the substrate and are shown as a 3D plot. In both cases, the length of the trajectories clearly decreases with increasing salt concentration demonstrating that the propulsion speed of the particles in bulk and close to the substrate is strongly dependent on the presence of salt.
\\
\\
\textbf{2D MSD analysis.}  Next, we make this assessment quantitative by extracting the effective diffusion coefficients $D$ and velocities $v$ from the particle trajectories by calculating and fitting the mean-squared displacement (MSD). In experiments close to the substrate, the particles are expected to have a fixed orientation with respect to the substrate~\cite{Das2015, Simmchen2016}. $D$ and $v$ are then obtained by fitting the experimentally measured MSDs with
\begin{align}
\label{eq:msd2d} \langle r^{2} \rangle_{\mathrm{2D}} &= 4D\tau + v^{2} \tau^{2} , 
\end{align}
which captures the short-time behavior in 2D~\cite{Bechinger2016,Howse2007}. This only holds, when the direction of self-propulsion is parallel along the wall and height fluctuations are small. 
\\
\\
\textbf{Parallel swimming assumption.} To test if this requirement is fulfilled, we use a microscopy setup where the stage and the objective can be tilted (see S.I. for more details). We record the trajectories of our active PS-swimmers in 0.1~wt\% \ce{H2O2} at different tilt angles $\theta$ and track the particle positions using trackpy~\cite{trackpy}. We then calculate the MSD and fit it with Eq.~\ref{eq:msd2d} up to lag times much smaller than the rotational timescale (0.5~s $\ll$ 70~s). 
We observe that $D$ remains constant with $\theta$, cf. S.I., which implies that the particles swim at a constant distance from the substrate, irrespective of the tilt angle. 
\\
\\
This also suggests a constant orientation of the swimmer with respect to the substrate. That is, a coupling between the substrate and the swimmer orientation that is stronger than any gravitational torque from the heavier metal cap. Because swimmers of different speeds have the same diffusion constant (Fig.~\ref{fig4}), we can assume that the swimmers propulsion force is aligned parallel to the substrate. If this would not be the case, a change in propulsion force, i.e. swimming speed, would have a component directed toward the substrate. This would cause a change in height and hence in the diffusion constant. Since we do not observe this, the assumption of parallel swimming holds. Therefore, in general we can compare the MSDs of different particles, as well as 2D and 3D experiments.
\\
\\
\textbf{3D MSD analysis} To analyze our bulk measurements, we need to take into account the contribution from sedimentation to isolate the effects from the activity. The obtained 3D-trajectories are therefore $z$-corrected for sedimentation by adding a linearly increasing height to the measured $z$-coordinate, $z(t)$, as $z_\text{new}(t) = z(t)+tv_\text{sedim}$. Here, $v_\text{sedim}>0$ is the obtained average sedimentation speed for passive particles, i.e. the same Pt/Pd half-coated PS particles in water, but in absence of fuel. The director of active particles that are moving downwards slower than $v_\text{sedim}$ or upwards is pointing upwards and their corrected speed in $z$ is thus positive. Likewise, the director of active particles moving downwards faster than $v_\text{sedim}$ is pointing downwards leading to a negative speed. For $v_\text{sedim}$, we find that all passive particles sediment with a similar speed of $0.7$~$\pm$~$0.1$~$\mu$ms$^{-1}$. This speed we extract from a total of 36 $z$-trajectories obtained using 9 passive particles by employing a linear fit, see S.I. for details. After taking into account sedimentation, we can fit the MSD for swimmers far away from the substrate. Then, the particles diffuse in 3D with
\begin{equation}
\label{eq:msd3d} \langle r^{2} \rangle_\mathrm{3D} = 6D\tau + v^2\tau^2 ,
\end{equation} where $\tau$ is the lag time while $D$ and $v$ again are fit constants. For both 2D and 3D experiments, lag times up to 1~s are fitted. For some trajectories, large displacements in $z$ between consecutive frames were observed for a specific height, which seems to indicate an inconsistency in the LUT of that particle. This could result in a positive offset of the MSD, reducing the quality of the fit. MSDs with an offset are therefore filtered out by only selecting MSDs where the difference between the first point, smallest lag time, and the fit is smaller than 50\% of the fit. See S.I. for details and for MSDs with their fits.
\\
\\
\textbf{Height-dependent MSD analysis.} Before measuring and comparing particle speeds at the substrate and in bulk, we first use the three-dimensional trajectories to determine the height at which the particles can be considered to move undisturbed in the bulk of the solution. To access this information, we divide the trajectory of a single particle into bins according to its distance from the substrate and calculate the MSD for every bin. The $z$-coordinate of the substrate is set equal to the initial $z$-value, as all trajectories started with the particle on the substrate. Over a total trajectory during which the acoustic tweezers were switched on multiple times, the $z$-coordinate of the substrate was noticed to change slightly. Therefore, we chose the first bin that includes the substrate to range from $0.75~\mathrm{\mu m}$ above to $0.75~\mathrm{\mu m}$ below the initial substrate height. In this bin, the data is treated as essentially 2D and the MSD is fitted with Eq.~\ref{eq:msd2d}. For all other bins, a bin size of $6~\mathrm{\mu m}$ for experiments in 0.1~wt\% \ce{H2O2} and 8~$\mathrm{\mu m}$ in 0.5~wt\% \ce{H2O2} is chosen, which balances a higher resolution in height (number of bins) with good statistics in each bin (bin size). The MSD for data of each bin is then fitted with Eq.~\ref{eq:msd3d} which corresponds to particles diffusing in 3D to obtain the fit constants $D$ and $v$, see Fig.~\ref{fig3}. Frames during the lifting of the particle with the acoustic tweezers and 1 second before and after that are not used in our calculations. Furthermore, only sufficiently long trajectories where the fit values $D$ and $v$ have an standard deviation smaller than 10\% are used (see S.I. for fits). All MSDs for the height-dependent data are fitted up to a lag time of $0.3~\mathrm{s}$. Finally, we normalize the diffusion constant with the one expected in the bulk as obtained from the Stokes-Einstein relation $D_{\mathrm{Stokes-Einstein}}=(k_{\mathrm{B}} T)/ (6\pi\eta r )$, where $T=296$~K, $\eta=0.9321$~mPas (water)~\cite{Huber2009}, and $r=2.25$~$\mu$m is used.
\\
\\
\textbf{Diffusion profile.} The evolution of $D(h)$ normalized over the bulk expectation value for $D_\mathrm{bulk}$ is shown in Fig.~\ref{fig3}a. We find that starting from the surface where $h/R \approx 0$, $D(h)/D_\mathrm{bulk}$ first sharply increases, before flattening out and approaching the bulk value. Despite being an active particle, the normalized values for the diffusion coefficient of the particle follow the hydrodynamic model derived by Faxén~\cite{Hilding1921}. This captures the height-dependent diffusion $D(h)$ of a \textit{passive}, spherical particle above a no-slip wall. Our results therefore confirm that it is possible to extract the particle-wall separation distance from the MSD of active particles, as was previously introduced by Ketzetzi~\textit{et al.}~\cite{Ketzetzi2020a}.

At 0.5~wt\% \ce{H2O2}, activity dominated already for small lag times, which made accurate extraction of the diffusive contribution difficult. That is, if the particle moves at a speed of 4 $\mathrm{\mu ms^{-1}}$ and has a diffusion constant of 0.1 $\mathrm{\mu m^2s^{-1}}$, at a frame rate of 33 fps, the activity term $v^2\tau^2$ is of the same order as  the diffusive term $6D\tau$ at the smallest lag time of 0.03~s, i.e. 0.014 $\mathrm{\mu m^2s^2}$ and 0.018 $\mathrm{\mu m^2s^2}$, respectively. For small lag times, a diffusion constant can be obtained, see S.I., however, these datapoints are sensitive to noise which results in the earlier described positive offset of the MSD at small lag times. Therefore, for the bulk experiment at 0.5~wt\% \ce{H2O2} without salt, the diffusion constant is not fitted, but instead set to $D_\text{Stokes-Einstein}$.
\\
\\
\textbf{Velocity profile.} For $v(h)$, we first look at the particles with a higher activity, i.e. those suspended in 0.5~wt\% \ce{H2O2} (see Fig.~\ref{fig3}b). The velocity first increases starting from ca. 2~$\mathrm{\mu ms^{-1}}$ close to the substrate before reaching a constant value of ca. 3.5~$\mathrm{\mu ms^{-1}}$ at heights corresponding to approximately 5 particle radii. However, when suspended in a lower fuel concentration of 0.1~wt\% \ce{H2O2} (see Fig.~\ref{fig3}b), the particles propel at a nearly constant velocity of about 1~$\mathrm{\mu ms^{-1}}$ at all heights. We conclude that the diffusion coefficient as well as the particle velocity at higher activity are sensitive to hydrodynamic effects which increase as the particle moves closer to the wall. This hydrodynamic influence of the substrate needs to be taken into account when comparing near-substrate and bulk behaviour of the active particles. Based on these observations, we define the bulk regime as starting from 5 particle radii above the substrate. Henceforth, bulk values for $D$ and $v$ are obtained from MSDs (of parts of the trajectories) at least 5$R$ away from the substrate, see SI for details.
\\
\\
\textbf{Effect of salt addition.} Having established the range above which we can consider our measurements to be in bulk, we now examine how salt affects the behaviour of active particles in this regime. 
The diffusion constants and velocities resulting from the analysis of $\langle r^{2} \rangle_\mathrm{2D}$ and $\langle r^{2} \rangle_\mathrm{3D}$ for experiments done in 0.1~wt\% \ce{H2O2} are shown in Fig.~\ref{fig4}a and b. For all salt concentrations, the diffusion constant obtained for swimmers far away from the substrate matches the value from the Stokes-Einstein relation to within the standard error of the mean. $D_\text{Stokes-Einstein}=(k_\text{B}T )/ (6\pi\eta r)$ resulting in $0.103$~$\mu$m$^2$s$^{-1}$ for particles with $r=$ 2.25~$\mu$m, $T=296$~K and $\eta=0.9321$~mPas~\cite{Huber2009} (water) as indicated by the black dashed line in Fig.~\ref{fig4}a. Close to the surface, the diffusion constant decreases to approximately half of the bulk value. The average of diffusion constants up to $c=0.5$~mM at the substrate is about 50\% of the bulk value, which corresponds to swimming heights of 0.38~$\mu$m using the interpolation of the models from Goldman and Faxén~\cite{Ketzetzi2020a}. This is in line with earlier observations of a swimming height of approximately 300 nm above the substrate~\cite{Ketzetzi2020a}. The decrease can predominantly be attributed to increased friction resulting from hydrodynamics.
\\
\\
Interestingly, we observe with increasing salt concentration that the velocities both close to the substrate and in bulk show a drastic decrease (see Fig.~\ref{fig4}b). Already for 0.1 mM NaCl the velocities decrease significantly and drop further to being close to zero at 10 mM in 0.1~wt\% \ce{H2O2}. While this decrease is in line with previous measurements of catalytically active particles on substrates~\cite{Brown2014,Ketzetzi2020a}, in bulk this significant reduction has not been observed before. The decrease of velocity in bulk suggests that ionic species affect the particle activity that originates from particle surface effects only. The similarity of the bulk and substrate velocities furthermore suggests that wall-effects acting on the particle speed are small in the presence of salt. The average velocity near the substrate seems to be a bit lower than the one in bulk, except for $c_\text{NaCl}=0$~mM where the errorbar is larger.
For 10 mM NaCl, insufficient data could be collected due to active particles sticking to the substrate which is why this data point was excluded in Fig.~\ref{fig4}a and b.
\\
\\
\textbf{Faster swimmers.} Next, we investigate if the aforementioned velocity differences are enhanced at higher particle speeds, that is for a system with 0.5~wt\% \ce{H2O2}. At this higher fuel concentration, the diffusion constant for bulk-swimmers still matches the value expected from the Stokes-Einstein relation (dashed line, see  Fig.~\ref{fig4}c) and again decreases close to the wall due to hydrodynamics. Note that the bulk value of $D$ for $c_\text{NaCl}=0$~mM is not shown because of the dominance of the velocity term for particles moving at these speeds, as explained earlier. The averaged diffusion constants close to the wall obtained for experiments in different fuel and salt concentrations up to 0.5 mM are very similar, being (0.054~$\pm$~0.007)~$\mu$m$^2$s$^{-1}$ and (0.058~$\pm$~0.023)~$\mu$m$^2$s$^{-1}$ in 0.1~wt\% \ce{H2O2} and 0.5~wt\% \ce{H2O2}, respectively. This indicates that the particle-wall separation distance is equal at both fuel concentrations, in line with previous work~\cite{Ketzetzi2020a}. 
\\
\\
For swimmers in the higher \ce{H2O2} concentration, we measure higher velocities. For any $c_\text{NaCl} > 0$~mM, particles move at similar speeds in bulk and close to the substrate, where again a small speed difference is observed, as in the lower fuel concentration. However, a notably large yet statistically significant velocity difference is observed between bulk and near-surface swimmers for 0 mM NaCl (see Fig.~\ref{fig4}d). We will provide possible explanations shortly.
\\
\\
\textbf{Speed difference substrate and bulk.} By comparing Figs.~\ref{fig4}b and d, we observe that the particle speed is very similar for bulk and substrate, with the exception of a single datapoint for the higher activity experiment at $c_\text{NaCl}=0$ (panel d). This difference could imply a surface effect that vanishes upon adding salt. However, at lower activity and in the absence of salt ($c_\text{NaCl}=0$ in panel b), the difference in particle speed between bulk and substrate is within the error. If there would be a surface effect that vanishes upon adding salt, it seems to be absent at low activities. 
\\
\\
Another explanation for the observed speed difference could be hydrodynamic effects. 
We base this insight on Fig.~\ref{fig3}b, where the velocity profile seems to follow the same scaling as the hydrodynamic Faxén model, as indicated by the dashed line. If the swimmer speed follows a scaling relation with height, the absolute difference between substrate and bulk should be most pronounced at higher activity, since experimental noise might overrule small differences at lower activity. In Figs.~\ref{fig4}b and d, higher fuel concentration and lower ionic concentration leads to faster particles, as well as a larger absolute difference in speed between substrate and bulk, which suggests coupling between the wall and the swimmer's speed. Notably, the coupling between swimmer and wall already strongly impacts the particle orientation and height, well before it affects the speed.
\\
\begin{figure}[t!]
\centering
  \includegraphics[width=0.5\textwidth]{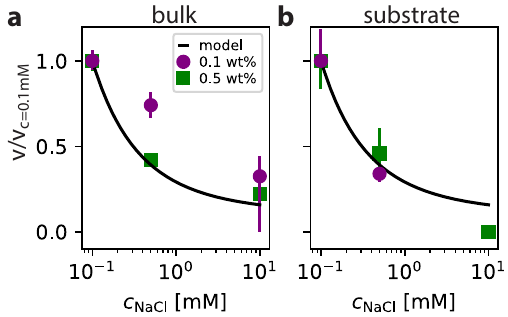}
  \caption{\textit{Relative speed reduction comparison.} The velocities relative to the velocity at $c_\text{NaCl}=0.1$~mM as function of $c_\text{NaCl}$ in bulk (a) and at the substrate (b) for two different fuel concentrations compared with a electrokinetic theory outlined in Ref.~\citenum{Brown2017}. Plotted points are median values and errorbars represent first and third quartiles.}
  \label{fig5}
\end{figure}
\\
\textbf{Implications for the propulsion mechanism.} We have seen that the effect of salt on the speed is equally present in the bulk as it is on the surface. This is a necessary step to gain a better understanding of the self-propulsion mechanism. The observation of a reduced swim velocity is consistent with the electrokinetic theory laid out in Refs.~\citenum{Brown2014, Brown2017}. The salt concentration used in the experiment spans two orders of magnitude, i.e. we have $\kappa a$ ranging from $\approx 2$ to $20$, where $a$ is the radius and $\kappa$ the inverse (Debye) screening  length\footnote[2]{For the purposes of this paper, we define $\kappa^{2} = 2 e^{2} c / ( \epsilon k_{\mathrm{B}} T )$ with $e$ the elementary charge, $\epsilon$ the dielectric constant of the medium, $k_{\mathrm{B}}$ Boltzmann's constant, and $T$ the temperature. In our calculations we assumed $\epsilon \approx 80 \epsilon_{0}$ with $\epsilon_{0}$ the vacuum permittivity and a temperature of $T \approx 300$~K.}. This lies outside the power-law scaling regime of the effective Henry's function for self-electrophoresis, $F_{\mathrm{H}}(\kappa a)$, as defined in Ref.~\citenum{Brown2017}. This function governs how the self-propulsion speed of a particle decreases as a function of decreasing $\kappa a$, when the only effect of salt on the particle speed is reducing the Debye screening length. Taking the speed at $c_\text{NaCl} = 0.1$~mM as a reference point, we predict a speed decrease by a factor of $\approx 0.4$ by increasing the salt concentration to $c_\text{NaCl} = 0.5$~mM, and a decrease by a factor of $\approx 0.15$ by increasing to $c_\text{NaCl} = 10$~mM, as shown in Fig.~\ref{fig5}. This appears to hold in bulk for higher activity $0.5$~wt\% \ce{H2O2} with 0.41 and 0.22 for the speed ratios at $c_\text{NaCl}$ values of respectively 0.5 mM and 10 mM relative to 0.1 mM, suggesting that our data aligns with the theoretical prediction, as shown in Fig. \ref{fig5}a.\\ 

However, the agreement is less good when looking at the bulk data for $0.1$~wt\% \ce{H2O2}, where these speed ratios are 0.75 and 0.32 and the data could fit more closely a logarithmic trend. This mismatch may imply that salt has a secondary effect at low fuel concentration, or another mechanism is at play. Looking at the substrate data shown in Fig. \ref{fig5}b, the points at $c_\text{NaCl}=0.5$~mM for both fuel concentrations match quite well with the theory, with 0.34 and 0.44 for fuel concentrations of 0.1~wt\% and 0.5~wt\%, respectively. The datapoint at $c_\text{NaCl}=10$~mM is missing for the 0.1~wt\% experiment as mentioned earlier. For the 0.5~wt\% experiment, the activity is so low that the obtained speed from fitting is equal to zero. This makes it hard to assess the quality of the trend. However, what is noticeable is that the trend for 0.1~wt\% substrate data strongly departs from that found in bulk, suggesting that the propulsion mechanism might have a different dominant component under these conditions. This difference in trend could also be due to the coupling between the substrate and particle being affected by salt, similar to osmotic counterflows as proposed in earlier work~\cite{Ketzetzi2020a}. Additional experiments will need to be conducted before conclusions can be drawn about the specific self-propulsion mechanism (or mechanisms) that is (or are) at play in our catalytic self-propelled particles.
\\

In summary, we have successfully quantified the diffusion constant and speed of active particles as a function of the distance from a wall. From this measurement, we could define the bulk behavior as starting five particle radii away from the substrate. We used this to measure the speed of active particles in bulk and compared it to particles moving close to a wall. Our results show that salt lowers the particle speeds both in 2D and 3D. At low activity, the measured particle speeds at different salt concentrations in both cases are very similar which suggests that ionic wall-effects have little effect on the particle speed. However, at higher activity we observe a speed difference between substrate and bulk that gets bigger with faster moving particles. This points towards an activity-dependent wall-effect of which the velocity height profile follows a scaling similar to a hydrodynamic model.\\

Furthermore, the observed speed reduction with increasing ionic concentration follows a scaling predicted by an electrokinetic theory for both microswimmers near a substrate and particles with a high activity in bulk. Particles with a lower activity in bulk follow a different scaling, which suggests a different propulsion mechanism, but additional experiments are needed. Our experiments thus not only provide a better understanding of ionic effects on microswimmers in 3D, but also suggest a coupling between the wall and the swimming speed. 

\section*{Author Contributions}
\textbf{Julio Melio:} Conceptualization, Methodology, Formal analysis, Investigation, Writing - Original Draft, Visualization\\
\textbf{Solenn Riedel:} Conceptualization, Methodology, Investigation, Writing - Original Draft, Visualization\\
\textbf{Ali Azadbakht:} Conceptualization, Methodology, Investigation\\
\textbf{Silvana Caipa Cure:} Methodology, Investigation\\
\textbf{Tom M.J. Evers:} Investigation\\
\textbf{Mehrad Babaei:} Investigation\\
\textbf{Alireza Mashaghi:} Resources\\
\textbf{Joost de Graaf:} Conceptualization, Methodology, Formal analysis, Writing - Original Draft, Funding acquisition\\
\textbf{Daniela J. Kraft:} Conceptualization, Methodology, Resources, Writing - Original Draft, Supervision, Funding acquisition

\section*{Conflicts of interest}
There are no conflicts to declare.


\section*{Acknowledgements}
We gratefully acknowledge funding from the European Research Council (ERC) under the European Union’s Horizon 2020 research and innovation program (grant agreement no. 758383, D.J.K.), from the Netherlands Organization for Scientific Research (NWO) through Start-Up Grant 740.018.013 (J.d.G.) and VIDI grant 193.069 (D.J.K). 


\scriptsize{
\bibliography{main_text} 
\bibliographystyle{rsc} } 

\end{document}


\title{{\Large\bf Supplementary Information:\\
The motion of catalytically active colloids approaching a surface}}

\author{Julio Melio}
\affleidenexp
\author{Solenn Riedel}
\affleidenexp
\author{Ali Azadbakht}
\affleidenexp

\author{Silvana A. Caipa Cure}
\affleidensim
\author{Tom M.J. Evers}
\affleidensim
\author{Mehrad Babaei}
\affleidensim
\author{Alireza Mashagi}
\affleidensim
\author{Joost de Graaf}
\affutrecht
\author{Daniela J. Kraft}
\affleidenexp

\date{\today}

\maketitle
\section{Tilted stage microscopy}
\begin{figure*}[b!]
    \centering
    \includegraphics[width=0.75\textwidth]{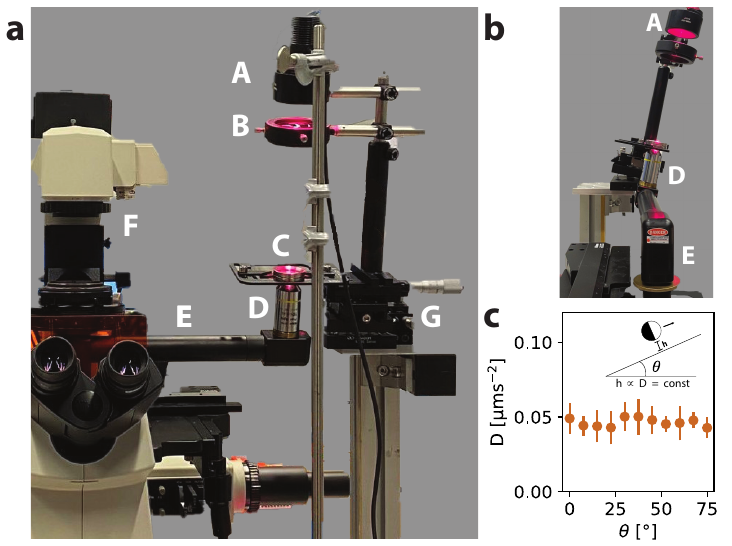}
    \caption{\textit{Tilted stage setup.} (a) Experimental setup where the stage and the objective of the microscope can be tilted. The microscope (F) is equipped with a rotating arm (E) on which a 10$\times$ objective (D) is mounted. The sample holder is placed on a stage that can be moved in $x$ and $y$ with the help of 2 linear stages and tilted to align with the objective using an angle plate (C and G). A 660 nm LED (A) and condensor (B) illuminate the sample. (b) Side view of the tilted setup. (c) Diffusion coefficient $D$ at different tilt angles $\theta$ resulting from the MSD analysis.}
    \label{fig:tilted_stage_setup}
\end{figure*}

The tilted stage microscopy setup used to determine the diffusion constant at different tilt angles is shown in Fig. \ref{fig:tilted_stage_setup}a. This setup consist of an inverted microscope (Nikon Ti-E) equipped with an arm (LSMTech InverterScope\textsuperscript{\textregistered}) onto which a 10$\times$ air objective was mounted. The arm is able to rotate the objective as can be seen in Fig. \ref{fig:tilted_stage_setup}b. Next to the microscope, a stage that could move in $x$ and $y$ and rotate with the objective was assembled using two linear stages and an angle plate. A 660~nm LED and a condensor lens were used to illuminate the sample. A diffuser was used to improve image quality (not shown).

Results of the diffusion constants at different tilt angles from MSD analysis are shown in Fig. \ref{fig:tilted_stage_setup}c. The diffusion constant stays constant (within error) which suggests that the particles keep swimming at the same distance from the substrate, regardless of the tilt angle.
\newpage
\section{Sedimentation speed}

\begin{figure}[hb!]
    \centering
    \includegraphics[width=0.4\linewidth]{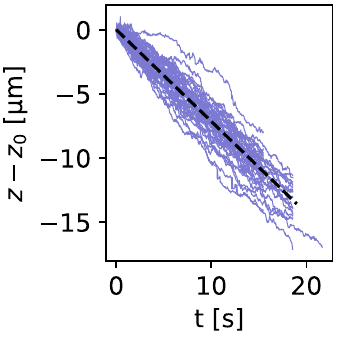}
    \caption{\textit{Sedimentation speed.} $z$-coordinate of trajectories of passive particles (PS-Pt/Pd particles in water) and the linear fit indicated with the black dashed line. The fit yields an average sedimentation speed of $0.7\pm0.1$~$\mu$ms$^{-1}$.}
    \label{fig:sedimentation_speed}
\end{figure}

The sedimentation speed was determined by analyzing in total 36 sedimentation occurrences of 9 different particles. The particles are the same as used in all experiments (4.5~$\mu$m PS, 5 nm~Pt/Pd) and these sedimentation experiments are done in water, which prevents active propulsion. In Fig. \ref{fig:sedimentation_speed}, the sedimentation trajectories are shown. For all trajectories, the first $z$-coordinate $z_0$ has been subtracted to make all trajectories start in the same point. The resulting data was then fitted with a linear function $z=at$, where $a$ is the fit constant corresponding to the sedimentation speed. From the fit, $a=-0.7\pm0.1$~$\mu$ms$^{-1}$ was extracted.

\newpage
\section{Additional information on Mean-squared displacement analysis}
\subsection{Height dependent mean squared displacements}
\begin{figure*}[htb!]
    \centering
    \includegraphics[width=\textwidth]{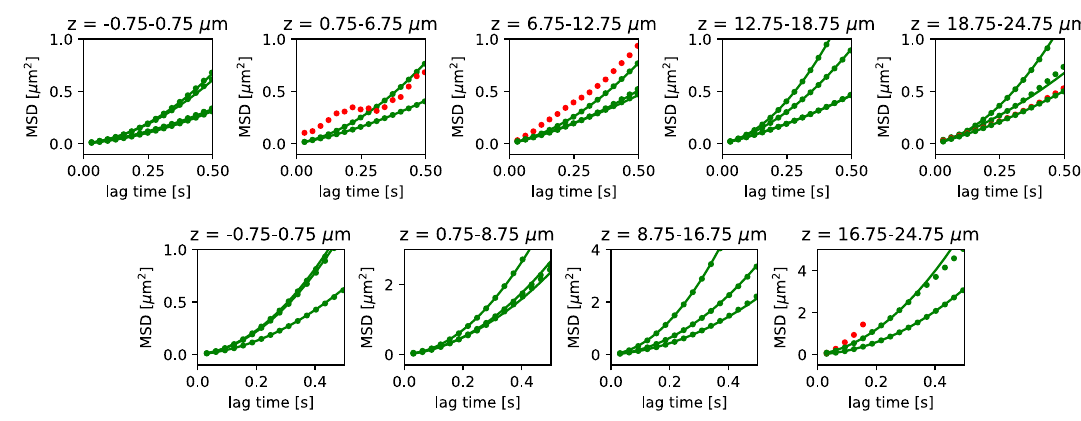}
    \caption{\textit{Height-dependent MSDs.} The Mean Squared Displacements (MSDs) for  parts of the trajectory that fall into different height bins with respect to the substrate. The height bin ranges are indicated in the titles of the subplots. The upper graphs are for experiments in 0.1 wt\% \ce{H2O2} and the lower graphs are for experiments in 0.5 wt\% \ce{H2O2}, both in absence of salt. MSDs that are sufficiently long, i.e. where the fitted $D$ and $v$ have a standard error smaller than 10\%, are shown in green, while the other MSDs are shown in red.}
    \label{fig:msds_height_dependent}
\end{figure*}
\newpage
\subsection{Fit quality and variance across sample}
\begin{figure*}[h!]
    \centering
    \includegraphics[width=0.9\textwidth]{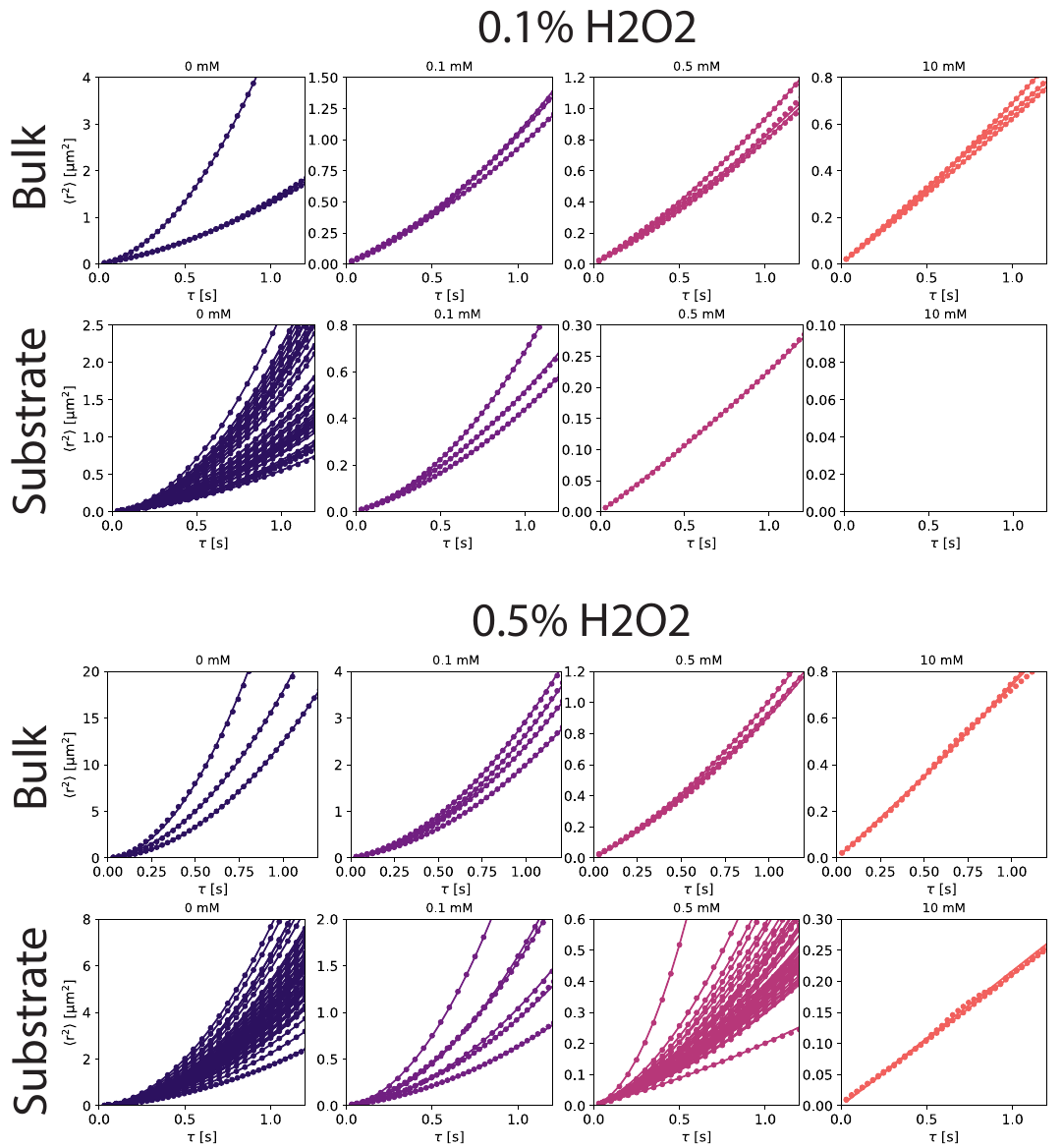}
    \caption{\textit{Mean-squared displacements and their fits.} For experiments in bulk and on a substrate, the measured MSDs from the trajectories are shown (indicated with points) and the fits are shown (indicated with a solid line). Lag times up to 1~s were considered for the fit.}
    \label{fig:msds_fit_quality}
\end{figure*}

In Fig. \ref{fig:msds_fit_quality}, the individual MSDs are shown with their corresponding fits. The  diffusion constants and velocities reported in the main text were extracted from this data. Both the good quality of the fit as well as the variance between particles can be observed in this representation. 

\newpage

\subsection{Temporary motion at certain height}
\begin{figure}[h!]
    \centering
    \includegraphics[width=0.6\textwidth]{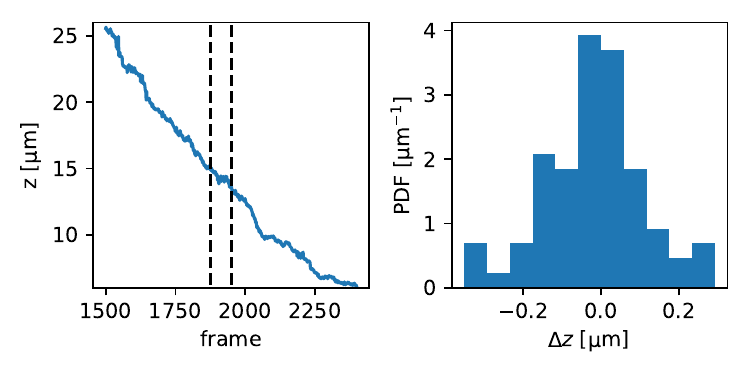}
    \caption{\textit{Larger displacements in $z$.} An example of a particle's $z$ trajectory where large displacements are observed at a specific height, leading to the shown displacement distribution that has small peaks around $\pm~0.3~\mathrm{\mu m}$.}
    \label{fig:jumps_in_z}
\end{figure}

Some $z$-trajectories contained large displacements at specific heights, of which an example is shown in Fig. \ref{fig:jumps_in_z}. The displacement distribution in the marked region is plotted on the right and shows two small peaks at low and high $\Delta z$. Because this is the displacement between consecutive frames, this will create an offset in the MSD. Therefore, only MSDs are selected where the difference between the fitted line and the first point, corresponding to the smallest lag time, is smaller than 50\% of the value of the fitted line.

\subsection{MSDs for different speeds and diffusion constants}
\begin{figure}[htb!]
    \centering
    \includegraphics[width=\textwidth]{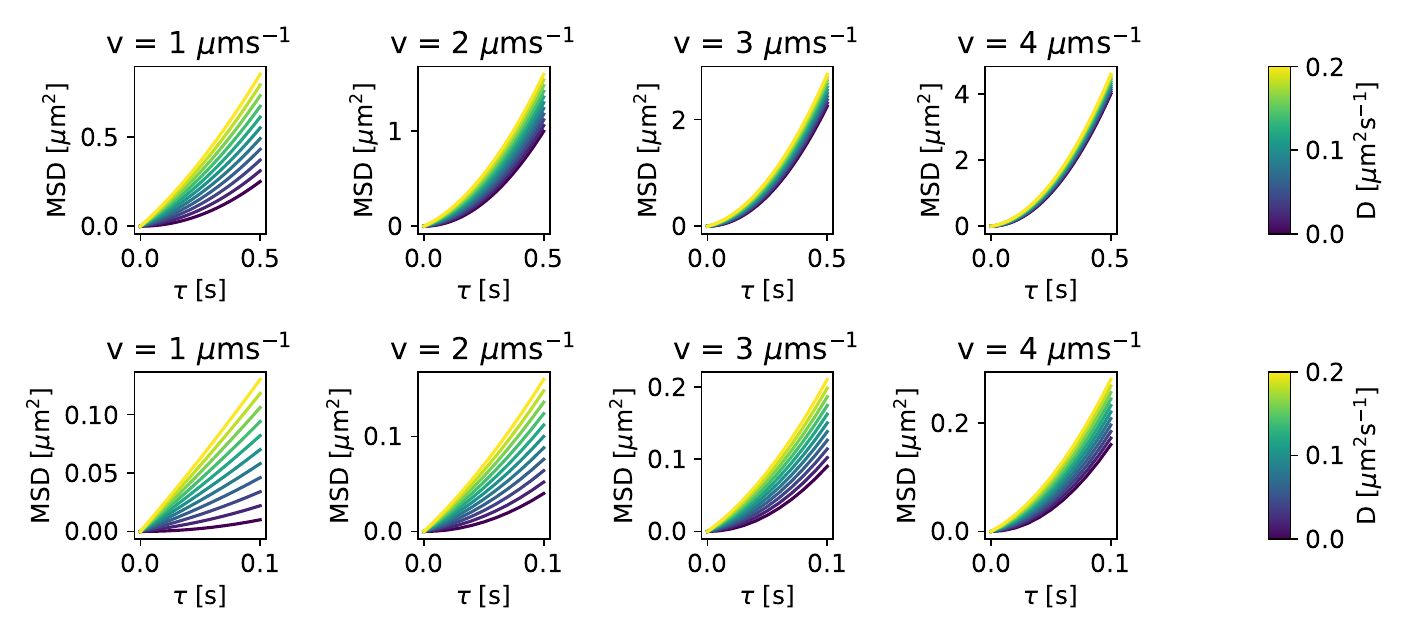}
    \caption{\textit{MSD for different speeds and diffusion constants} The mean squared displacement as calculated from $6D\tau+v^2\tau^2$ is plotted for different values of $D$ (represented by color) and different values of $v$ (indicated in the title) up to a lag time of $\tau=0.5$~s (top row) and $\tau=0.1$~s (bottom row).}
    \label{fig:time_where_6Dt_and_v2t2_are_equal}
\end{figure}